\documentclass{bmcart}

\usepackage[utf8]{inputenc} 

\usepackage{amssymb,amsfonts}
\usepackage{amsthm,amsmath}
\usepackage{todonotes}
\usepackage{mathtools}
\usepackage{algorithm}
\usepackage{arevmath} 
\usepackage{algpseudocode}
\usepackage{glossaries}
\makeglossaries
\newacronym{aes}{AES}{Advanced Encryption Standard}
\newacronym{aqc}{AQC}{Adiabatic Quantum Computing}
\newacronym{cnf}{CNF}{Conjunctive Normal Form}
\newacronym{crc}{CRC}{Cyclic Redundancy Check }
\newacronym{dnf}{DNF}{Disjunctive Normal Form}
\newacronym{mppk}{MPPK}{Multivariate Polynomial Public Key}
\newacronym{ecc}{ECC}{Elliptic Curve Cryptography}
\newacronym{eke}{EKE}{Encrypted Key Exchange}
\newacronym{pfp}{PFP}{Factoring Polynomial Problem}
\newacronym{grh}{GRH}{Generalized Riemann Hypothesis}
\newacronym{IND-CPA}{IND-CPA}{Indistinguishability Under Chosen-Plaintext Attack}
\newacronym{IND-CCA2}{IND-CCA2}{Indistinguishability Under Adaptive Chosen Ciphertext Attack}
\newacronym{kem}{KEM}{Key Encapsulation Mechanism}
\newacronym{lwe}{LWE}{Learning With Errors}
\newacronym{lwr}{LWR}{Learning With Rounding}
\newacronym{mdep}{MDEP}{Modular Diophantine Equation Problem}
\newacronym{mpkc}{MPKC}{Multivariate Public Key Cryptosystem}
\newacronym{nist}{NIST}{National Institute of Standards and Technology}
\newacronym{ntru}{NTRU}{Nth degree Truncated polynomial Ring Units}
\newacronym{pke}{PKE}{Public-Key encryption}
\newacronym{pki}{PKI}{Public Key Infrastructure}
\newacronym{pqc}{PQC}{Post-Quantum Cryptography}
\newacronym{rsa}{RSA}{Rivest-Shamir-Adleman}
\newacronym{sidh}{SIDH}{Super singular Isogeny Diffie-Hellman} 
\newacronym{svp}{SVP}{Shortest Vector Problem}
\newacronym{tls}{TLS}{Transport Layer Security}

\def\includegraphics{}

\startlocaldefs
\endlocaldefs

\begin{document}

\begin{frontmatter}

\begin{fmbox}
\dochead{Research}


\title{Benchmark Performance of  Homomorphic Polynomial Public Key Cryptography for Key Encapsulation and Digital Signature Schemes}


\author[
   addressref={aff1},                   
   corref={aff1},                       
   email={randy.kuang@quantropi.com}   
]{\inits{RK}\fnm{Randy} \snm{Kuang}}
\author[
   addressref={aff1},
   email={maria.perepechaenko@quantropi.com}
]{\inits{MP}\fnm{Maria} \snm{Perepechaenko}}
\author[
   addressref={aff1},
   email={dafu.lou@quantropi.com}
]{\inits{DL}\fnm{Dafu} \snm{Lou}}
\author[
   addressref={aff1},
   email={brinda.tank@quantropi.com}
]{\inits{BT}\fnm{Brinda} \snm{Tank}}



\address[id=aff1]{                           
  \orgname{Quantropi Inc.}, 
  \street{1545 Carling Ave, Suite 620},       
  \postcode{K1Z 8P9},                         
  \city{Ottawa},                              
  \cny{Canada}                                
}




\begin{artnotes}
\note[id=n1]{Equal contributor} 
\end{artnotes}

\end{fmbox}


\begin{abstractbox}

\begin{abstract} 
This paper presents a comprehensive benchmarking analysis of the Homomorphic Polynomial Public Key (HPPK) Key Encapsulation Mechanism (KEM) and Digital Signature (DS), recently introduced by Kuang et al. Departing from traditional cryptographic approaches, these schemes leverage the security of homomorphic symmetric encryption across two hidden rings without relying on NP-hard problems. HPPK can be considered a specialized variant of Multivariate Public Key Cryptography (MPKC), intricately associated with two vector spaces: the polynomial vector space for secret exchange and the multivariate vector space for randomized encapsulation.

Given the unique integration of asymmetric, symmetric, and homomorphic cryptography within HPPK, a meticulous examination of its performance metrics is imperative. This study focuses on a comprehensive benchmarking of HPPK KEM and DS, spanning key cryptographic operations, including key generation, encapsulation, decapsulation, signing, and verification. The results underscore the exceptional efficiency of HPPK, characterized by compact key sizes, cipher sizes, and signature sizes. The incorporation of symmetric encryption enhances overall performance. Key findings highlight the outstanding performance of HPPK KEM and DS across various security levels, emphasizing their superiority in critical cryptographic operations. This research positions HPPK as a promising and competitive solution for post-quantum cryptographic applications across diverse domains such as blockchain, digital currency, and Internet of Things (IoT) devices.
\end{abstract}




\end{abstractbox}
%

\end{frontmatter}




\section{Introduction}
In the dynamic realm of cryptography, the pursuit of robust and efficient cryptographic schemes has recently achieved remarkable progress, particularly with the groundbreaking innovations introduced by Kuang et al. These innovations include the Homomorphic Polynomial Public Key (HPPK) Key Encapsulation Mechanism (KEM) and Digital Signature (DS), both evolving from earlier stages like DPPK~\cite{kuang2021ACCC} and MPPK~\cite{mppkkem2022, mppk-quadractic-case}. Departing from conventional cryptographic paradigms, these schemes leverage homomorphic symmetric encryption across two concealed rings, offering a distinct avenue for ensuring security without resorting to NP-hard problems.

HPPK emerges as a specialized variant of Multivariate Public Key Cryptography (MPKC)~\cite{Ding2009mpkc}, intricately intertwining two vector spaces: a polynomial vector space for secret exchange and a multivariate vector space for randomized encapsulation. This departure from convention allows HPPK to seamlessly integrate asymmetric, symmetric, and homomorphic cryptography, presenting a novel paradigm with broad applications.

The security foundation of HPPK relies on symmetric encryption with a self-shared key, manifesting as two co-prime pairs $GCD(R_1, S_1)=1$ and $GCD(R_2, S_2)=1$ with $S_1$ and $S_2$ defining the rings. The symmetric encryption key, not shared with the encryptor and verifier, necessitates partial homomorphic properties like addition and scalar multiplication. These properties, inherent in modular multiplication encryption of polynomial coefficients, align with the polynomial public key structure. The self-shared key serves a dual purpose: first, encrypting the plaintext public key while preserving its mathematical structure, and subsequently, decrypting the received ciphertext of the secret through the asymmetric mechanism. The computational complexity of finding the pairs $(R_1, S_1)$ and $(R_2, S_2)$ for HPPK KEM is $\mathcal{O}(\eta (S_1^2 +S_2^2))\rightarrow\mathcal{O}(\eta 2^{2L})$, with $L=log_2S_1 =log_2S_2$ being the size of the rings and the constant $\eta < 1$ signifying reduced brute force searches due to the co-prime condition. Once the symmetric key is revealed, there is no additional computational difficulty in finding other private key elements.

In contrast, HPPK DS~\cite{HPPK-DS-kuang-2023-xchive} requires a transformative approach to eliminate the moduli $S_1$ and $S_2$ for signature verification. Utilizing the Barrett reduction algorithm for efficient modular multiplication, it transforms verification polynomials, establishing a non-linear relationship with the signature embedded coefficients. This characteristic significantly mitigates the potential for forged signatures.

The primary objective of this paper is to conduct a comprehensive benchmarking analysis of the performance of HPPK KEM and DS. Through a meticulous examination of key cryptographic operations, including key generation, encapsulation, decapsulation, signing, and verification, we aim to provide a thorough understanding of the schemes' efficiency. This evaluation extends to comparisons of key sizes, cipher sizes, and signature sizes, offering valuable insights into the practical viability of these schemes.

This study positions HPPK at the forefront of post-quantum cryptographic solutions, emphasizing its exceptional performance and adaptability across various security levels. Beyond theoretical considerations, the practical implications of HPPK extend to applications such as blockchain, digital currency, and Internet of Things (IoT) devices. Through this benchmarking analysis, we aim to contribute valuable insights into the evolving landscape of cryptographic schemes and their practical implications in contemporary security contexts.

\section{Related Works}

The domain of Post-Quantum Cryptography (PQC) encompasses a spectrum of standardized schemes identified by the National Institute of Standards and Technology (NIST). This overview provides a succinct summary of noteworthy schemes categorized based on their cryptographic foundations. For Key Encapsulation Mechanism (KEM), we have lattice-based Kyber~\cite{KYBER}, BIKE~\cite{BIKE}, HQC~\cite{HQC}, and code-based McEliece~\cite{McEliece1978}. Additionally, for Digital Signature (DS), the lattice-based Falcon~\cite{falcon}, Dilithium~\cite{dilithium}, and hash-based SPHINCS$^+$~\cite{sphincs} are highlighted.

In 2022, NIST took a significant stride by announcing its standardized algorithms~\cite{NIST8413}, endorsing Kyber for KEM and advancing McEliece, BIKE, and HQC into round 4. Concurrently, NTRU~\cite{Bernstein2018} and Saber~\cite{SABER} were excluded from further consideration, while new submissions for generic digital signature schemes were introduced~\cite{NIST8413}.

Lattice-based algorithms, such as Kyber, BIKE, HQC, and Falcon, typically rely on the Short-Vector Problem (SVP) as the cornerstone of their security. Code-based algorithms, exemplified by McEliece, hinge on the complexity of decoding random linear codes, providing post-quantum security. Hash-based algorithms, as seen in SPHINCS$^+$, are constructed on the security of one-way trapdoors in hash functions. These NP-hard problems form the foundation of security against the impending threat of quantum computing. In contrast, HPPK cryptography takes a distinctive approach by building on the security of symmetric encryption, offering a unique and innovative path in the landscape of post-quantum cryptographic solutions.

\section{Brief HPPK Cryptography}
\label{sec:hppk}
We present a succinct overview of HPPK cryptography, emphasizing the shared characteristics of the Key Encapsulation Mechanism (KEM) and Digital Signature (DS) schemes. Subsequently, we delve into the details of KEM and DS in separate subsections.

HPPK cryptography, as introduced by Kuang et al.~\cite{kuang2023-HPPK-KEM, HPPK-DS-kuang-2023-xchive}, starts from three polynomials: two univariate polynomials \(f(x)\) and \(h(x)\), where \(x\) signifies the secret, and one multivariate polynomial \(\beta(x, u_1, \dots, u_m)\) over the prime field \(\mathbb{F}_p\). The latter involves noise variables \(u_j\in \mathbb{F}_p\) for randomized encapsulations of the chosen secret \(x\) and signature verification for a given message. These polynomials follow general forms:

\begin{align*}
    &f(x) = f_0 + f_1 x + \dots + f_{\lambda} x^{\lambda} \\
    &h(x) = h_0 + h_1 x + \dots + h_{\lambda} x^{\lambda} \\
    &\beta(x, u_1, \dots, u_m) = \sum_{i=0}^n \beta_i(u_1, \dots, u_m) x^i = \sum_{i=0}^n \sum_{j=1}^m c_{ij}  x^iu_j
\end{align*}

Using polynomial multiplication, two product polynomials \( p(x, u_1, \dots, u_m)\) and \( q(x, u_1, \dots, u_m)\) are constructed, leading to public key coefficients \(p_{ij}\) and \(q_{ij}\). HPPK cryptography introduces the homomorphic operator \(\hat{\mathcal{E}}_{(R, S)}\) and its decryption counterpart. These operators can be applied to polynomials on their coefficients.

The HPPK KEM scheme involves creating two hidden rings marked by \({R_1, S_1}\) and \({R_2, S_2}\). The homomorphic operators \(\hat{\mathcal{E}}_{(R_1, S_1)}\) and \(\hat{\mathcal{E}}_{(R_2, S_2)}\) are applied to the coefficients of public polynomials \(P(.)\) and \(Q(.)\). The key pair consists of private keys \(R_1, S_1; R_2, S_2; f[\lambda +1], h[\lambda +1]\) and public keys \(P[n+\lambda +1][m]\) and \(Q[n+\lambda +1][m]\).

In the encryption process, an encrypter randomly chooses a secret \(x\) and noise variables \(u_1, \dots, u_m\). The resulting ciphertext, a tuple \(C=\{\overline{P}, \overline{Q}\}\), is sent to the decrypter. The decrypter calculates \(k\) based on the received ciphertext and solves for \(x\) using the univariate polynomial equation \(f(x) - kh(x) = 0 \bmod{p}\).

It's important to note that the benchmarking analysis considers specific choices for parameters, such as \(\lambda=1\), to optimize the performance of HPPK cryptography.

\subsection{HPPK KEM}\label{sec:keyPairConstrKEM}
Now the coefficients of public polynomials \(P(\cdot)\) and \(Q(\cdot)\) form the public key. Let's summarize the key pair:

\begin{itemize}
    \item Security parameters: \(\{n, \lambda, p\}\);
    \item \textbf{Private key}: \(R_1,S_1; R_2, S_2; f[\lambda +1], h[\lambda +1]\);
    \item \textbf{Public key}: \(P[n+\lambda +1][m]\) and \(Q[n+\lambda +1][m]\) 
\end{itemize}

Using the public key, an encrypter randomly chooses a secret \(x\) in \(\mathbb{F}_p\) to be encapsulated. The encryption also requires randomly chosen \(m\) values \(u_1, \dots, u_m \in \mathbb{F}_p\) for the noise variables, and then evaluates two polynomial values as follows:

\begin{equation}
    \begin{aligned}
       \overline{P} &= P(x, u_1, \dots, u_m) = \sum_{i=0}^{n+\lambda} \sum_{j=1}^m P_{ij} (u_jx^i \bmod{p}) \\
       \overline{Q} &= Q(x, u_1, \dots, u_m) = \sum_{i=0}^{n+\lambda} \sum_{j=1}^m Q_{ij} (u_jx^i  \bmod{p}) \\        
    \end{aligned}
\end{equation}

Then, the ciphertext is a tuple \(C=\{\overline{P}, \overline{Q}\}\) sent to the decrypter for the secret extraction.

The decrypter receives the ciphertext tuple \(C=\{ \overline{P}, \overline{Q}\}\). Then, the decrypter calculates:

\begin{equation}\label{eq:decrypt}
\begin{aligned}
 k &= \frac{R_1^{-1}[\overline{P} = P(x, u_1, \dots, u_m)] \bmod{S_1}}{R_2^{-1}[\overline{Q} = Q(x, u_1, \dots, u_m)] \bmod{S_2}} \bmod{p} \\
&= \frac{\beta(x,u_1, \dots, u_m)f(x)}{\beta(x,u_1, \dots, u_m)h(x)} \bmod{p} \\
&= \frac{f(x)}{h(x)}  \bmod{p}
\end{aligned}
\end{equation}

The value \(k\) in Eq.~(\ref{eq:decrypt}) is evaluated from the received ciphertext tuple. At this point, the decrypter needs to solve for \(x\) from the following univariate polynomial equation:

\begin{equation}\label{eq:decrypt2}
f(x) - kh(x) = 0 \bmod{p}.
\end{equation}

Recall that \(f(x)\) and \(h(x)\) are solvable polynomials of degree \(\lambda\). Eq.~\eqref{eq:decrypt2} can be solved with well-known radicals. Thanks to symmetric homomorphic encryption over hidden ring(s), the optimal choice for the order of polynomials \(f(x)\) and \(h(x)\) would be linear to avoid possible more than one root requiring an extra verification. Therefore, this benchmark uses \(\lambda=1\).

\subsection{HPPK DS}
Based on Eq.~\eqref{eq:decrypt}, we can perform a cross-multiplication and obtain the following equation:

\begin{align}\label{eq:verify}
    &[f(x) R_2^{-1} Q(x, \Vec{u}) \bmod{S_2}] \bmod{p} = [h(x) R_1^{-1} P(x, \Vec{u}) \bmod{S_1}] \bmod{p} \nonumber \\
    &\longrightarrow [F(x) Q(x, \Vec{u}) \bmod{S_2}] \bmod{p}=[H(x) P(x, \Vec{u}) \bmod{S_1}] \bmod{p}
\end{align}
where $\Vec{u}$ denotes the vector $(u_1, \dots, u_m)$ and Eq.~\eqref{eq:verify} behaves like a verification equation with signature elements defined as:

\begin{equation}\label{eq:signature}
    F(x) = f(x) R_2^{-1} \bmod{S_2} ; \quad H(x) = h(x) R_1^{-1}  \bmod{S_1}
\end{equation}
with $f(x)$ and $h(x)$ to be evaluated with $\bmod{\ p}$ then decrypted into rings $\mathbb{Z}_{S_2}$ and $\mathbb{Z}_{S_1}$ respectively. Considering the unknown moduli $S_1$ and $S_2$ in Eq.~\eqref{eq:verify}, the signature verifier could not perform the verification. We have to transform Eq.~\eqref{eq:verify} into a new form without $S_1$ and $S_2$. The Barrett reduction algorithm is applied for this transformation as described in the paper~\cite{HPPK-DS-kuang-2023-xchive}. In order to apply the Barrett reduction algorithm for modular multiplications, let's rewrite Eq'~\eqref{eq:verify} to the following form by expanding polynomials $P(\cdot)$ and $Q(\cdot)$:

\begin{align}\label{eq:verify1}
      &\sum_{j=1}^m \sum_{i=0}^{n+\lambda} V_{ij}(F) x^iu_j \bmod{p} 
     = \sum_{j=1}^m \sum_{i=0}^{n+\lambda} U_{ij}(H) x^iu_j \bmod{p} \nonumber \\
     &\longrightarrow V(F, x, u_1, ..., u_m) = U(H, x, u_1, ..., u_m) \bmod{p}
\end{align}
with polynomial coefficients $V_{ij}(F)$ and $U_{ij}(H)$ defined as:

\begin{equation} \label{eq:uv}
\begin{aligned}
    U_{ij}(H) &= [H*P_{ij} \bmod{S_1}] \bmod{p}  \\
    V_{ij}(F) &= [F*Q_{ij} \bmod{S_2}] \bmod{p}.
\end{aligned}
\end{equation}
where $F=F(x)$ and $H=H(x)$ are signature elements with $x\longleftarrow HASH(M)$ representing the hash code of a signing message $M$. Using the Barrett reduction algorithm, we can transform coefficients of verification polynomials in Eq.~\eqref{eq:uv} into the following equations by multiplying a randomly chosen $\beta\in\mathbb{F}_p$ and then taking $\bmod{\ p}$:

\begin{align}\label{eq:uv1}
    U_{ij}(H) &=  H*p'_{ij} -s_1\lfloor \frac{H\mu_{ij}}{R} \rfloor \bmod{p}\nonumber \\
    V_{ij}(F) &= F*q'_{ij} -s_2\lfloor \frac{F\nu_{ij}}{R} \rfloor \bmod{p}.
\end{align}
with 

\begin{equation}\label{eq:publicDS}
 \begin{aligned}
     &s_1 = \beta S_1 \bmod{p}  \\
     &s_2 = \beta S_2 \bmod{p}  \\
     &p'_{ij} = \beta P_{ij} \bmod{p}  \\
     &q'_{ij} = \beta Q_{ij} \bmod{p}  \\
     &\mu_{ij} = \lfloor \frac{RP_{ij}}{S_1} \rfloor   \\
     &\nu_{ij} = \lfloor \frac{RQ_{ij}}{S_2} \rfloor
 \end{aligned}
 \end{equation}
to be the public key for signature verification. In Eq.~\eqref{eq:publicDS}, $R=2^K$ is the Barrett parameter as a security parameter with $K >> L$. HPPK signing is described by Eq.~\eqref{eq:signature} and verification by Eq.~\eqref{eq:verify1}.

\section{Security Summary}
The detailed security analysis is available in HPPK KEM for dual hidden rings\cite{kuang2023-HPPK-KEM} and for a single hidden ring\cite{hppk-f1000-2023}. In this paper, we provide a concise summary of the conclusions. The security of HPPK KEM primarily stems from symmetric homomorphic encryption over hidden rings using the self-shared keys ${R_1, S_1; R_2, S_2}$ for dual rings and ${R_1, S_1; R_2, S_2=S_1}$ for a single ring. Once the symmetric key is discovered, all other private key elements can be easily unveiled without computational difficulty.

Discovering the symmetric encryption key involves knowing $R_1$ and $S_1$ over a hidden ring $1$ and $R_2$ and $S_2$ over a hidden ring $2$, achieved through random guessing with a complexity of $\mathcal{O}(S_1^2)$ and $\mathcal{O}(S_2^2)$, respectively. For each guessed $S_1$, the attacker must bruteforce $R_1$ and test if $R_1$ is coprime with $S_1$. For each found coprime pair $(R_1, S_1)$, the attacker must use it to decrypt the public key $P[n+\lambda+1][m]$ and verify if all decrypted $P[n+\lambda+1][m]\in \mathbb{F}_p$. If not, the process is repeated. This is why the complexity is $\mathcal{O}(S_1^2)$. The same process applies to ring $2$. Therefore, the overall complexity is $\mathcal{O}(S_1^2 + S_2^2)=\mathcal{O}(2*2^{2L})$ with $|S_2|_2=|S_1|_2=L$, for dual hidden rings. For a single hidden ring, the complexity could be just $\mathcal{O}(2^{2L})$. We can safely include a factor $\eta<1$ for the complexity $\mathcal{O}(\eta2^{2L})$ to consider the some effective way for coprime pair searches. 

Considering ciphertext-only attacks, the complexity for the secret recovery attack is $\mathcal{O}(p^{m-1})$, requiring the number of noise variables to be more than one for a non-deterministic secret recovery attack, as we have two equations established from the public polynomials $P(.)$ with ciphertext $\overline{P}$ and $Q(.)$ with $\overline{Q}$.

For NIST security levels I, III, and V, we choose the bit length $L=2*|p|_2 +8$ bits. Then the overall complexity is $\mathcal{O}(2^{4\log_2 p+16})$. Table~\ref{tab:Config} illustrates different configurations for the three NIST security levels. It is evident that the two variants OHR and THR of HPPK KEM only differ in private key, as there is no $S_2$ for HPPK-OHR. Therefore, their performance should be similar for encryption and decryption, but HPPK-OHR might be slightly more efficient for key generation due to the smaller private key size. This paper will focus on benchmarking the performance of HPPK-OHR for KEM.
\begin{table}[ht]
\caption{Configurations of HPPK KEM over One-Hidden-Ring or OHR and Two-Hidden-Rings or THR, for different NIST security levels, given as a quadruple ($\log p, n, \lambda, m$).}
\begin{center}
\begin{tabular}{cccc}
\hline
&\multicolumn{3}{c}{\textbf{Security}}\\
 & \textbf{\textbf{Level I}}& \textbf{\textbf{Level III}}& \textbf{\textbf{Level V}}\\
\hline
Entropy against key recovery (bit)& 144              & 208                 &  272      \\
Configurations       & (32, 1, 1, 2) & (48, 1, 1, 2), & (64, 1, 1, 2)\\
($PK, SK:OHR|THR, CT$) in Bytes & (108, 43$|$52, 224) & (156, 63$|$76, 240) & (204, 83$|$100, 208) \\
Configurations & (32, 1, 1, 3) & (48, 1, 1, 3), & (64, 1, 1, 3)\\
($PK, SK:OHR|THR, CT$) in Bytes & (162, 43$|$52, 224) & (234, 63$|$76, 240) & (306, 83$|$100, 208) \\

\hline
\end{tabular}
\label{tab:Config}
\end{center}
\end{table}

Within the HPPK DS framework, Kuang et al. introduced a key recovery attack leveraging specific public key values, denoted as $s_1$ and $s_2$~\cite{HPPK-DS-kuang-2023-xchive}. The initial method for determining moduli $S_1$ and $S_2$ exhibited a computational complexity of $\mathcal{O}(S_1S_2/p)=\mathcal{O}(2^{2L}/p)$. This paper presents an optimized key recovery attack directly utilizing public key elements $\mu_{ij}$ and $\nu_{ij}$, defined as:

\begin{equation}
\begin{aligned}
&\mu_{ij} = \lfloor \frac{R \cdot p_{ij}}{S_1} \rfloor \longrightarrow p_{ij} = \lceil \frac{S_1 \cdot \mu_{ij}}{R} \rceil\\
&\nu_{ij} = \lfloor \frac{R \cdot q_{ij}}{S_2} \rfloor \longrightarrow q_{ij} = \lceil \frac{S_2 \cdot \nu_{ij}}{R} \rceil
\end{aligned}
\end{equation}

This streamlined attack involves iteratively searching for $S_1$ within the range $2^{L-1}$ to $2^L$, calculating $p_{ij}$ using the known public key $\mu_{ij}$, and recalculating $\mu_{ij}$ with $S_1$ and $p_{ij}$. If the recomputed $\mu_{ij}$ matches the public key $\mu_{ij}$, the attacker deterministically identifies the private values $S_1$ and $p_{ij}$. The computational complexity of this approach is $\mathcal{O}(2^{L-1})$. A similar procedure applies to $S_2$ and $q_{ij}$, resulting in a total complexity of $\mathcal{O}(2^L)$.

Additionally, the remaining private key elements can be effortlessly determined by intercepting genuine signatures with the known values of $p_{ij}$, $q_{ij}$, $S_1$, and $S_2$. This optimized key recovery attack exhibits a complexity of $\mathcal{O}(2^{L})$, showcasing significantly enhanced computational efficiency compared to the previous approach with a complexity of $\mathcal{O}(2^{2L}/p)$.

Table~\ref{Tab:keyandsignsizes} demonstrates that different configurations do not affect private key and signature sizes, only influencing public key sizes. For $m=2$, as the Barrett parameter $k$ decreases from $K=2L$ to $K=L+32$, the public key size significantly decreases from 544B to 376B for level I, from 792B to 528B for level III, and from 1040B to 680B for level V, respectively. On the other hand, the public key size is reduced by almost $50\%$ when $m$ is changed from 2 to 1, as shown in the 4th configuration for all three security levels.

It is evident that the hash algorithms SHA-256, SHA-384, and SHA-512 are recommended for security levels I, III, and V, respectively. The signature size will be 32B for level I, 48B for level III, and 64B for level V. With selected primes as shown in Table~\ref{Tab:keyandsignsizes}, hash codes will be segmented into four segments, each being 8B for level I, 12B for level 3, and 16B for level V. For $m=1$, there are four quadratic equations producing four sets of roots, creating $2^4$ possible forked hash codes associated with $2^4$ possible messages. This allows some forged messages to pass verification. Taking an example for the SHA-256 hash algorithm, the collision rate is generally $\frac{1}{2^{256}}$, so the probability of a forged signature is at a level of $\frac{2^4}{2^{256}}$ for $m=1$ and $\frac{2^3}{2^{256}}$ for $m=2$. In practical terms, an attacker would not gain any meaningful advantage for a forged signature attack by reducing $m$ from 2 to 1. However, the public key size would be dramatically reduced. In this benchmarking, we present the HPPK DS performance of key generartion, signing, and verifying with $m=1$.

\begin{table}[h!]
    \caption{The key and signature sizes in bytes, as provided by the HPPK DS scheme for the proposed parameter sets, are determined based on the optimal complexity of $\mathcal{O}(2^{L})$. In this context, we choose the Barrett parameter $R$ to be $32$ bits longer than $S_1/S_2$, and the hidden ring size is set to be $L=2\times|p|_2 + 16$. All data is presented in bytes. The configuration is defined as $(n, \lambda, m, L, log_2R). $}
    \centering
{\begin{tabular}{cccccccc} 
 \hline
Security & p& Configuration &Entropy (bits)  & $PK$& $SK$ & $Sig$ & $Hash$\\ 
 \hline
 I   &$2^{64}-59$ & (1,1,2, 144, 288)& 144& 544 & 104 & 144 & SHA-256 \\
     &$2^{64}-59$ & (1,1,2, 144, 208)& 144& 424 & 104 & 144 & SHA-256 \\
     &$2^{64}-59$ & (1,1,2, 144, 176)& 144& 376 & 104 & 144 & SHA-256 \\
     &$2^{64}-59$ & (1,1,1, 144, 208)& 144& 220 & 104 & 144 & SHA-256 \\
     &$2^{64}-59$ & (1,1,1, 144, 176)& 144& \textbf{196} & 104 & 144 & SHA-256 \\
 III &$2^{96}-17$ &(1,1,2, 208, 416)& 208&  792 &152 & 208 & SHA-384\\ 
     &$2^{96}-17$ &(1,1,2, 208, 272)& 208&  576 &152 & 208 & SHA-384\\ 
     &$2^{96}-17$ &(1,1,2, 208, 240)& 208&  528 &152 & 208 & SHA-384\\
     &$2^{96}-17$ &(1,1,1, 208, 272)& 208&  300 &152 & 208 & SHA-384\\
     &$2^{96}-17$ &(1,1,1, 208, 240)& 208&  \textbf{276} &152 & 208 & SHA-384\\
 V   &$2^{128}-159$&(1,1,2, 272, 544)& 272&  1040 & 200& 272 & SHA-512\\
     &$2^{128}-159$&(1,1,2, 272, 336)& 272&  728 & 200& 272 & SHA-512\\
     &$2^{128}-159$&(1,1,2, 272, 304)& 272&  680 & 200& 272 & SHA-512\\
     &$2^{128}-159$&(1,1,1, 272, 336)& 272&  380 & 200& 272 & SHA-512\\
     &$2^{128}-159$&(1,1,1, 272, 304)& 272&  \textbf{356} & 200& 272 & SHA-512\\
 \hline
\end{tabular}}
\label{Tab:keyandsignsizes}
\end{table}

\section{Benchmarking Results}\label{sec:performance}
In this section, we will demonstrate the performance of HPPK KEM in Subsection ~\ref{sec:kem-ben} and HPPK DS in subsection~\ref{sec:ds-ben}. We used the SUPERCOP benchmarking tool~\cite{SUPERCOP}. All schemes have been configured to achieve the NIST security levels~I, III, and~V. The three levels correspond to the difficulty of breaking 128, 192, and 256-bit \gls*{aes}.  We ran SUPERCOP on a 16-core Intel\textregistered Core\texttrademark i7-10700 CPU at 2.90 GHz system.

\subsection{HPPK KEM}\label{sec:kem-ben}

The fundamental operations of the HPPK Key Encapsulation Mechanism (KEM) scheme, including key generation, encapsulation, and decapsulation, are outlined in Algorithms~\ref{alg:keygen-kem}, Algorithm~\ref{alg:encrypt}, and Algorithm~\ref{alg:decrypt} correspondingly. Table~\ref{tab:keysize-kem} provides a thorough comparison of key sizes and ciphertext sizes for the HPPK KEM, juxtaposed with NIST-standardized Kyber and round 4 candidates McEliece, BIKE~\cite{BIKE}, and HQC~\cite{HQC}.

At Security Level I, which mandates a secret key size of 32 bytes or more, cryptographic schemes display varied sizes for public keys, private keys, and ciphertexts. McEliece exhibits large key sizes, boasting a public key of 261,120 bytes and a private key of 6,492 bytes. In contrast, Kyber, BIKE, and HQC present relatively smaller key sizes, ranging from a few hundred to a few thousand bytes. Notably, HPPK-(32,1,1,2) and HPPK-(32,1,1,3) introduce compact key structures, featuring a public key of 108 or 162 bytes, a private key ranging from 43 to 52 bytes, and a ciphertext of 224 bytes. The primary impact on the public key size stems from an increase in the number of noise variables.

Security Level III imposes elevated security requirements, targeting 192 bits of entropy, resulting in larger key sizes for most cryptographic schemes. McEliece maintains substantial key sizes for resilience, and Kyber, BIKE, and HQC witness incremental size adjustments. Conversely, HPPK-(48,1,1,2) and HPPK-(48,1,1,3) efficiently adapt to the increased security level, offering smaller key sizes. They feature a public key of 156 or 234 bytes, a private key ranging from 63 to 76 bytes, and a ciphertext of 240 bytes.

Security Level V, demanding the highest security standards, leads to larger key sizes across cryptographic schemes. McEliece continues to exhibit substantial key sizes for robust security, while Kyber, BIKE, and HQC scale up, emphasizing their adaptability to increased security requirements. Significantly, HPPK-(64,1,1,2) and HPPK-(64,1,1,3) remain efficient at this high security level, featuring a public key of 204 or 306 bytes, a private key ranging from 83 to 100 bytes, and a ciphertext of 208 bytes. These reduced sizes underscore the effectiveness of HPPK KEM in achieving a balance between security and efficiency at the highest security level.
\begin{algorithm}
\caption{Key generation of HPPK KEM.}\label{alg:keygen-kem}
\begin{algorithmic}[1]
\Procedure{KeyGen}{$\lambda$, $n$, $m$, $p$}
   
    \For{$(i=0; i\le \lambda; i++)$}  \Comment{loop i}
         \State $f_i\gets Random() \mod p$ \Comment{generate $f(x)$}
         \State $h_i\gets Random() \mod p$ \Comment{generate $h(x)$}

    \EndFor  \label{poly loop}

    \For{$(i=0; i\le n; i++)$}  \Comment{loop i}
         \For{$(j=0; j< m; j++)$}  \Comment{loop j} 
            \State $c_{ij} \gets Random() \mod p$ \Comment{generate $\beta(x, u_1, \dots, u_m)$}
         \EndFor  
    \EndFor  \label{poly loop}

     \State 
    \For{$(i=0; i\le n+\lambda; i++)$}  \Comment{Evaluate public key $P(.), Q(.)$}
        \For{$(j=0; j< m; j++)$}
            \For{$(s=0; s< i; s++)$}
                \State $p_{ij} \gets f_s c_{(i-s)j}$  
                \State $Q_{ij} \gets h_s c_{(i-s)j}$
            \EndFor
        \EndFor
    \EndFor 
    
    \State
    \State $\ell \gets 2  \log_2 p + 8 $    \Comment{Make the hidden field 8 bits larger than doubled prime field}
    \State $S_1 \gets Random(\ell) $  \Comment{Generate the hidden ring $\mathbb{Z}/S_1\mathbb{Z}$}
    \State  $R_1 \gets Random(\ell) \bmod{S_1}$
    \While{$gcd(R_1, S_1) \not= 1$}
        \State  $R_1 \gets Random(\ell) \bmod{S_1}$
    \EndWhile

    \State
    \State $S_2 \gets Random(\ell) $  \Comment{Generate the hidden ring $\mathbb{Z}/S_2\mathbb{Z}$}
    \State  $R_2 \gets Random(\ell) \bmod{S_2}$
    \While{$gcd(R_2, S_2) \not= 1$}
        \State  $R_2 \gets Random(\ell) \bmod{S_2}$
    \EndWhile           \Comment{Complete private key $SK: f[], h[], R_1, S_1, R_2, S_2$}

     \State 
    \For{$(i=0; i\le n+\lambda; i++)$}  \Comment{Evaluate public key $PK: P(.), Q(.)$}
        \For{$(j=0; j< m; j++)$}
            \For{$(s=0; s< i; s++)$}
                \State $P_{ij} \gets R_1 * P_{ij} \bmod{S_1}$  
                \State $Q_{ij} \gets R_2 * Q_{ij} \bmod{S_2}$
            \EndFor
        \EndFor
    \EndFor         \Comment{Complete public key $PK: P[n+\lambda +1][m], Q[n+\lambda +1][m]$}
\EndProcedure
\State \textbf{return} $SK, PK$  \Comment{Return private key SK and public key}
\end{algorithmic}
\end{algorithm}

\begin{algorithm}
\caption{HPPK encapsulation. $P$ and $Q$ are $(n+\lambda+1)\times m$ matrices with security parameters $p$, $n$, $\lambda$.}\label{alg:encrypt}
\begin{algorithmic}[1]
\Procedure{encapsulation}{$P$, $Q$}
    \For{$(j=1; j\le m; j++)$}
        \State $u_j \gets Random()  \ \bmod{p}$  
    \EndFor
    \State 
    \State  $\overline{P} \gets 0$
    \State  $\overline{Q} \gets 0$
    
    \For{$(i=0; i\le n+\lambda+1; i++)$}  \Comment{Evaluate  $\overline{P}, \overline{Q}$}
        \For{$(j=1; j\le m; j++)$}
            \State $\overline{P}  \gets P_{ij} (u_j s^i \ \bmod{p})$  
            \State $\overline{Q}  \gets Q_{ij} (u_j s^i \ \bmod{p})$
        \EndFor
    \EndFor  
    
\EndProcedure
\State \textbf{return} $\overline{P}, \overline{Q} $  \Comment{Return ciphertext}
\end{algorithmic}
\end{algorithm}

\begin{algorithm}
\caption{HPPK decapsulation. Inputs include  ciphertext $C= \{\overline{P}, \overline{Q}\}$, prime $p$, private key $SK=\{f[\lambda+1], h[\lambda+1]\}, R_1, R_2, S_1, S_2$}\label{alg:decrypt}
\begin{algorithmic}[1]
\Procedure{decapsulation}{$C, \overline{P}, \overline{Q}$}
    \State $\overline{P} \gets \left(\frac{\overline{P}}{R_1} \ \bmod{S_1} \right) \ \bmod{p}$  \Comment{homomorphic decryption of $\overline{P}$}
    \State $\overline{Q} \gets \left(\frac{\overline{Q}}{R_2} \ \bmod{S_2} \right) \ \bmod{p}$ \Comment{homomorphic decryption of $\overline{Q}$}

    \State $k \gets \frac{\overline{P}}{\overline{Q}}   \bmod{p}$ 
    
    \State
    \State  $s \gets solve \ f(s) - kh(s) = 0 \ \bmod(p)$ \Comment{use radical to solve the roots}
\EndProcedure
\State \textbf{return} $s$
\end{algorithmic}
\end{algorithm}

\begin{table}[htbp]
\caption{This table compares public key, private key, and ciphertext sizes for HPPK KEM, NIST-standardized Kyber, and round 4 candidates McEliece, BIKE, and HQC. HPPK KEM offers two modes—single and double hidden rings—impacting private key size denoted as $SK_1$ and $SK_2: SK_1/SK_2$. The analysis provides insights into key size efficiency and security trade-offs across different security levels.}
\begin{center}
\begin{tabular}{ccccc}
\hline
\textbf{Crypto}&\multicolumn{4}{c}{\textbf{Size (Bytes)}} \\
\cline{2-5} 
\textbf{system} & \textbf{\textit{Public key (PK)}}& \textbf{\textit{Private key(SK)}}& \textbf{\textit{Ciphertext}} & \textbf{\textit{Secret}}\\
\hline
    &\multicolumn{4}{c}{\textbf{Security Level I}} \\
McEliece & 261,120 & 6,492 & 128 & 32\\
Kyber & 800	& 1,632 & 768 & 32 \\
BIKE~\cite{BIKE} & 1,541        & 281       & 1,573 & 32\\
HQC~\cite{HQC}  &2,249&56&4,497&64\\
HPPK-(32,1,1,2) & 108 & 43/52 & 224 & 32 \\
HPPK-(32,1,1,3) & 162 & 43/52 & 224 & 32 \\
   &\multicolumn{4}{c}{\textbf{Security Level III}} \\
   McEliece & 524,160 & 13,608 & 188 & 32\\
Kyber & 1,184	& 2,400 & 1,088 & 32 \\
BIKE~\cite{BIKE} & 3,083 & 419 & 3,115 &32\\
HQC~\cite{HQC}  &4,522&64&9,042&64\\
HPPK-(48,1,1,2) & 156 & 63/76 & 240 & 32 \\
HPPK-(48,1,1,3) & 234 & 63/76 & 240 & 32 \\
   &\multicolumn{4}{c}{\textbf{Security Level V}} \\
   McEliece & 1,044,992 & 13,932 & 240 & 32\\
Kyber & 1,568	& 3,168 & 1,568 & 32 \\
BIKE~\cite{BIKE} & 5,122 & 581 & 5,904\\
HQC~\cite{HQC}  &7,245&72&14,485&64\\
HPPK-(64,1,1,2) & 204 & 83/100 & 208 & 32 \\
HPPK-(64,1,1,3) & 306 & 83/100 & 208 & 32 \\
\hline
\end{tabular}
\label{tab:keysize-kem}
\end{center}
\end{table}
The comparative performance analysis of key generation, encapsulation, and decapsulation for different cryptographic schemes at Security Levels I, III, and V is presented in Table~\ref{tab:kem}. It is crucial to note that the performance metrics for BIKE are derived from their AVX2 implementation due to susceptibility to side-channel attacks in their reference implementation~\cite{BIKE}.

At Security Level I, McEliece exhibits a relatively high key generation time of 152,424,455 clock cycles, reflecting its design and larger key sizes. In contrast, Kyber, BIKE(AVX2), and HQC offer more efficient key generation processes, with Kyber leading in clock cycles. For encapsulation at Security Level I, HPPK-(32,1,1,2) and HPPK-(32,1,1,3) outperform other schemes with 25,963 and 65,776 clock cycles, respectively, showcasing their efficiency. McEliece, BIKE(AVX2), and Kyber demonstrate comparable performance for encapsulation, while HQC is relatively slower. For decapsulation at Security Level I, HPPK-(32,1,1,2) and HPPK-(32,1,1,3) stand out, requiring only 63 kilocycles, making them standout performers. McEliece is the slowest, followed by BIKE and HQC, with Kyber being the second fastest.

At Security Level III, McEliece experiences increased key generation times, demonstrating its resilience but highlighting scalability challenges. BIKE(AVX2) is the second slowest, followed by HQC. Kyber maintains fast key generation, while HPPK-(48,1,1,2) and HPPK-(48,1,1,3) variants exhibit efficient key generation, adapting well to heightened security requirements. For encapsulation at Security Level III, HPPK-(48,1,1,2) excels with 30,452 clock cycles, showcasing its efficiency. McEliece and Kyber have comparable performance, while HQC is relatively slower. For decapsulation at Security Level III, HPPK-(48,1,1,2) and HPPK-(48,1,1,3) continue their efficient performance, requiring only 54 kilocycles, one third of the cycles of the fastest scheme Kyber. McEliece, BIKE(AVX2), and HQC are the slowest, second slowest, and third slowest schemes, respectively.

At Security Level V, McEliece experiences substantial key generation times. Kyber and HQC demonstrate increased cycles compared to lower security levels. HPPK-(64,1,1,2) and HPPK-(64,1,1,3) variants stand out with efficient key generation, requiring about 20 kilocycles. For encapsulation at Security Level V, HQC is the slowest scheme, while McEliece and Kyber demonstrate comparable performance. HPPK-(64,1,1,2) maintains efficiency with 16,941 clock cycles, less than 10

The performance trends of HPPK KEM from Security Level I to V for encapsulation and decapsulation showcase intriguing characteristics. Performance at lower security levels takes more clock cycles due to the smaller field size, necessitating more segments for encapsulation and decapsulation for the given NIST-required minimum 32 bytes of the secret. However, at Security Level III, where the field size is 48 bits, the secret is 26 bytes long and segmented into 6 segments, enabling superior performance. This characteristic allows a single Security Level V to be considered for both Security Level I and Security Level III, providing better encapsulation and decapsulation performance, albeit with slightly larger key and ciphertext sizes, as shown in Table~\ref{tab:keysize-kem}. Considering key sizes and performance, opting for the HPPK KEM scheme with two noise variables appears optimal, offering sizes comparable to RSA-2048.

\begin{table}[htbp]
\caption{Comparison of key generation, encapsulation, and decapsulation performance for HPPK KEM is illustrated  with NIST standardized Kyber and round 4 candidates McEliece, BIKE, and HQC. Performance data for BIKE and HQC are cited from your NIST submission specifications and for McEliece and Kyber are directly computed from the the same SUPERCOP tool as HPPK KEM schemes. }
\begin{center}
\begin{tabular}{ccccc}
\hline
\textbf{Crypto}&\multicolumn{3}{c}{\textbf{Performance (Clock cycles)}} \\
\cline{2-4} 
\textbf{system} & \textbf{\textit{KeyGen}}& \textbf{\textit{Encapsulation}}& \textbf{Decapsulation} \\
\hline
    &\multicolumn{3}{c}{\textbf{Security Level I}} \\
McEliece        & 152,424,455   & 108,741   & 45,122,734 \\
Kyber           & 72,403        &95,466     &	117,406 \\
BIKE(AVX2)~\cite{BIKE}& 589,000       &	97,000 &	1,135,000 \\
HQC~\cite{HQC}  & 187,000       &	419,000  &	833,000 \\
HPPK-(32,1,1,2) & 12,665        &   25,963  &   63,365\\
HPPK-(32,1,1,3) & 20,098        &   65,776  &   63,729\\ \\
    &\multicolumn{3}{c}{\textbf{Security Level III}} \\
McEliece        & 509,364,485   &172,538    & 93,121,707 \\
Kyber           & 115,654       & 140,376   &	166,062 \\
BIKE(AVX2)~\cite{BIKE}& 1,823,000     &223,000    &	3,887,000\\
HQC~\cite{HQC}  & 422,000       &946,000	&  1,662,000\\
HPPK-(48,1,1,2) & 18,327        &30,452     &  54,653\\
HPPK-(48,1,1,3) & 22,831        &40,495     &  53,164\\ \\
    &\multicolumn{3}{c}{\textbf{Security Level V}} \\ 
McEliece        & 1,127,581,201 &263,169&	179,917,368 \\
Kyber           & 177,818       &205,504	  &237,484 \\
HQC~\cite{HQC}  & 830,000       &1,833,000	&  3,343,000\\
HPPK-(64,1,1,2) & 19,416        &16,941     &  29,026\\
HPPK-(64,1,1,3) & 26,931        &22,307     &  28,176\\
\hline
\end{tabular}
\label{tab:kem}
\end{center}
\end{table}

\subsection{HPPK DS}\label{sec:ds-ben} 
The procedural details of HPPK DS are explicated in Algorithm~\ref{alg:keygenerationDS} for key generation, Algorithm~\ref{alg:sign} for the signing process, and Algorithm~\ref{alg:verify} for the verification of signatures. A comprehensive overview of key sizes and signature sizes for HPPK DS is presented in Table~\ref{tab:DS_sizes}, facilitating a comparative analysis with well-established NIST-standardized algorithms, including the lattice-based Dilithium~\cite{dilithium} and Falcon~\cite{falcon}, as well as the hash-based SPHINCS$^+$~\cite{sphincs}.

\begin{algorithm}
\caption{Key generation of HPPK DS.}\label{alg:keygenerationDS}
\begin{algorithmic}[1]
\Procedure{KeyGenDS}{$\lambda$, $n$, $m$, $p$}
   
    \For{$(i=0; i\le \lambda; i++)$}  \Comment{loop i}
         \State $f_i\gets Random() \mod p$ \Comment{generate $f(x)$}
         \State $h_i\gets Random() \mod p$ \Comment{generate $h(x)$}

    \EndFor  \label{poly loop}

    \For{$(i=0; i\le n; i++)$}  \Comment{loop i}
         \For{$(j=0; j< m; j++)$}  \Comment{loop j} 
            \State $c_{ij} \gets Random() \mod p$ \Comment{generate $\beta(x, u_1, \dots, u_m)$}
         \EndFor  
    \EndFor  \label{poly loop}

     \State 
    \For{$(i=0; i\le n+\lambda; i++)$}  \Comment{Evaluate public key $P(.), Q(.)$}
        \For{$(j=0; j< m; j++)$}
            \For{$(s=0; s< i; s++)$}
                \State $p_{ij} \gets f_s c_{(i-s)j}$  
                \State $Q_{ij} \gets h_s c_{(i-s)j}$
            \EndFor
        \EndFor
    \EndFor 
    
    \State
    \State $\ell \gets 2  \log_2 p + 16 $    \Comment{Make the hidden field 16 bits larger than doubled prime field}
    \State $S_1 \gets Random(\ell) $  \Comment{Generate the hidden ring $\mathbb{Z}/S_1\mathbb{Z}$}
    \State  $R_1 \gets Random(\ell) \bmod{S_1}$
    \While{$gcd(R_1, S_1) \not= 1$}
        \State  $R_1 \gets Random(\ell) \bmod{S_1}$
    \EndWhile

    \State
    \State $S_2 \gets Random(\ell) $  \Comment{Generate the hidden ring $\mathbb{Z}/S_2\mathbb{Z}$}
    \State  $R_2 \gets Random(\ell) \bmod{S_2}$
    \While{$gcd(R_2, S_2) \not= 1$}
        \State  $R_2 \gets Random(\ell) \bmod{S_2}$
    \EndWhile           \Comment{Complete the private key $SK: f[], h[], R_1, S_1, R_2, S_2$ }

     \State 
     \State  $\beta \gets Random() \bmod{p}$ \Comment{Randomly choose $\beta$ to provide extra protection of $S_1, S_2$}
    \For{$(i=0; i\le n+\lambda; i++)$}  \Comment{Evaluate public key $P(.), Q(.)$}
        \For{$(j=0; j< m; j++)$}
            \For{$(s=0; s< i; s++)$}
                \State $P_{ij} \gets R_1 * P_{ij} \bmod{S_1}$  
                \State $Q_{ij} \gets R_2 * Q_{ij} \bmod{S_2}$
                \State $p'_{} \gets \beta*P_{ij} \bmod{p}$   \Comment{DS public key}
                \State $q'_{ij} \gets \beta*Q_{ij} \bmod{p}$
                \State $\mu_{ij} \gets \lfloor \frac{P_{ij} >> K}{S_1} \rfloor$ \Comment{K: the Barrett parameter}
                \State $\nu_{ij} \gets \lfloor\frac{Q_{ij} >> K}{S_2} \rfloor$
                \State $s_1 \gets \beta *S_1 \bmod{p}$
                \State $s_2 \gets \beta *S_2 \bmod{p}$
            \EndFor
        \EndFor
    \EndFor     \Comment{Complete the public key $PK: p'[], q'[], \mu[][], \nu[][], s_1, s_2$ }
\EndProcedure
\State \textbf{return} $SK, PK$  \Comment{Return public key}
\end{algorithmic}
\end{algorithm}

\begin{algorithm}
\caption{HPPK DS signing. $P$ and $Q$ are $N\times m$ matrices with security parameters $p$, $n$, $\lambda$.}\label{alg:sign}
\begin{algorithmic}[1]
\Procedure{encrypt}{$f[]$, $h[], R_1, S_1, R_2, S_2, m, x$}
    
    \State 
    \State  $F \gets 0$ 
    \State  $H \gets 0$
    
    \For{$(i=0; i\le n+\lambda+1; i++)$}  \Comment{Evaluate  signature elements $F$ and $H$}
        \For{$(j=1; j\le m; j++)$}
            \State $F  \gets f_{i} * x^i \ \bmod{p})$  
            \State $F  \gets f_{i} * x^i \ \bmod{p})$
        \EndFor
    \EndFor
    \State $F \gets R_2^{-1} * F \bmod{S_2}$   \Comment{map to hidden rings}
    \State $H \gets R_1^{-1} * H \bmod{S_1}$ 
    
\EndProcedure
\State \textbf{return} $F, H $  \Comment{Return signature}
\end{algorithmic}
\end{algorithm}

\begin{algorithm}
\caption{HPPK DS verification. Inputs include  signature $F, H, PK_v$, prime $p$}\label{alg:verify}
\begin{algorithmic}[1]
\Procedure{Verify}{$F, H, PK_v$}
    \State $Result \gets true$
     \For{$(j=1; j\le m; j++)$}
        \State $LHS \gets 0$    \Comment{LHS: the verification polynomial value on the left hand side.}
        \State $RHS \gets 0$    \Comment{RHS: the verification polynomial value on the right hand side.}
        \For{$(i=0; i\le n+\lambda+1; i++)$}  \Comment{Evaluate  scoefficients $U_{ij}$ and $V_{ij}$}
            \State $U_{ij}  \gets H * p'_{ij} -s_1*\lfloor \frac{H*\mu_{ij}}{R} \rfloor\ \bmod{p})$   \Comment{Barrett's parameter $R=2^k$ }
            \State $V_{ij}  \gets F * q'_{ij} -s_1*\lfloor \frac{F*\nu_{ij}}{R} \rfloor\ \bmod{p})$
            \State $LHS \gets LHS + U_{ij}*x^i \ \bmod{p}$
            \State $RHS \gets RHS + V_{ij}*x^i \ \bmod{p}$
        \EndFor
        \If { $( LHS != RHS )$} 
            \State $Result \gets false$
            \State $break;$
        \EndIf
    \EndFor
    
\EndProcedure
\State \textbf{return} $Result$
\end{algorithmic}
\end{algorithm}

Table~\ref{tab:DS_sizes} provides a visual representation of the comparisons in key size and signature size among various cryptographic schemes. Notably, SPHINCS$^+$ stands out for its smallest key sizes, not exceeding 128 bytes for all three security levels. However, it demonstrates the largest signature sizes, measuring 7,856 bytes for Security Level I, 16,224 bytes for Security Level III, and 29,792 bytes for Security Level V.

In the domain of lattice-based schemes, Dilithium exhibits larger sizes for public key, private key, and signature compared to Falcon across all three security levels, with sizes generally measured in kilobytes. In contrast, HPPK DS demonstrates a remarkable optimization, presenting more compact sizes: 220 bytes for the public key, 104 bytes for the private key, and 144 bytes for the signature at Security Level I; 300 bytes, 152 bytes, and 208 bytes, respectively, at Security Level III; and 380 bytes, 200 bytes, and 272 bytes, respectively, at Security Level V.

\begin{table}[htbp]
\caption{Comparison of public key size, private key size, and signature size for HPPK DS is illustrated  with NIST standardized Dilithium and finalists. $L$ is selected to be $2*|p| + 16$ and the Barrett parameter $K=L+64$ bits is used in this benchmarking. Optimized HPPK DS refers to a configuration $(64, 1, 1, 1)$ with $L=168$ bits.}
\begin{center}
\begin{tabular}{cccc}
\hline
\textbf{Crypto}&\multicolumn{3}{c}{\textbf{Size (Bytes)}} \\
\cline{2-4} 
\textbf{system} & \textbf{\textit{Public key}}& \textbf{\textit{Private key}}& \textbf{\textit{Signature}} \\
\hline
    &\multicolumn{3}{c}{\textbf{Security Level I}} \\
Dilithium 2 &1312 &     & 2420 \\
Falcon512 & 897	& 1281 & 690  \\
SPHINCS$^+$-128s~\cite{sphincs} & 32 & 64 & 7856 \\
HPPK-(64,1,1,1) & 220 & 104 & 144  \\ \\
   &\multicolumn{3}{c}{\textbf{Security Level III}} \\
Dilithium 3 & 1592	& 4016  & 3293  \\
SPHINCS$^+$-192s~\cite{sphincs} & 32 & 64 & 16224 \\
HPPK-(96,1,1,1) & 300 & 152 & 208  \\ \\
   &\multicolumn{3}{c}{\textbf{Security Level V}} \\
Dilithium 5 & 2592	&  4880 & 4595  \\
Falcon1024 & 1793	& 2305 & 1330  \\
SPHINCS$^+$-256s~\cite{sphincs}&64 & 128 & 29792 \\
HPPK-(128,1,1,1) & 380 & 200 & 272  \\
\hline
\end{tabular}
\label{tab:DS_sizes}
\end{center}
\end{table}

Table~\ref{tab:DS_sign} provides a comprehensive performance comparison for key generation, signing, and verifying across various cryptographic schemes, including HPPK DS, Dilithium, Falcon, and SPHINCS$^+$. The performance metrics are detailed for Security Levels I, III, and V, with the HPPK DS configuration considered involving $n=\lambda=m=1$ and utilizing the Barrett parameter $K=L+64$ bits.

At Security Level I, Dilithium 2 demonstrates competitive key generation performance at about 300 kilocycles, while HPPK-(64,1,1,1) showcases a significantly lower number of cycles at about 26 kilocycles. Falcon512 key generation exhibits comparatively higher values with more than 38 million cycles. Notably, SPHINCS$^+$-128s requires a substantial number of cycles, with 358 million cycles for key generation. For signing operations, HPPK-(64,1,1,1) stands out as the most efficient, requiring only 12,510 cycles, whereas Dilithium 2 and Falcon512 demonstrate higher cycle counts, exceeding one million cycles. SPHINCS$^+$-128s exhibits the highest number of cycles, over 2.7 billion, reflecting its hash-based nature. In terms of verification, HPPK-(64,1,1,1) continues to perform exceptionally well, requiring only 18,349 cycles. Dilithium 2 and Falcon512, although higher than HPPK DS, demonstrate reasonable verification performance. SPHINCS$^+$-128s again exhibits the highest cycle count due to its hash-based approach.

Moving to Security Level III, Dilithium 3 showcases competitive key generation performance~\cite{dilithium}. Our Supercop results for Dilithium 3 are about 2x faster than their performance for key generation, signing, and verification~\cite{dilithium}. While SPHINCS$^+$-192s requires a significant number of cycles, more than 524 million cycles for key generation. HPPK DS showcases a reasonable increase in clock cycles due to the bigger finite field size, ranging from 25 to 35 kilocycles. For signing operations, HPPK-(64,1,1,1) maintains efficient performance with 14,382 cycles, outperforming Dilithium 3 with about 1 million cycles. SPHINCS$^+$-192s exhibits a higher cycle count, almost doubling its cycles from security level I, reaching 5 billion cycles, reflective of its hash-based structure. Signature verification sees HPPK-(64,1,1,1) once again demonstrating efficiency with 21,145 cycles, slightly increasing from its cycles at security level I. Dilithium 3 would be 15x slower than HPPK. SPHINCS$^+$-192s requires a substantial number of cycles due to its hash-based nature.

At Security Level V, Dilithium 5~\cite{dilithium} exhibits competitive key generation performance with 819 kilocycles, while Falcon1024 requires a significantly higher number of cycles, over 100 million cycles. Our Supercop results for Dilithium 5 are again about 2x faster than their performance for key generation, signing, and verification~\cite{dilithium}. HPPK-(128,1,1,1) maintains its efficient key generation with 42 kilocycles, again slightly increasing its cycles from security level III. SPHINCS$^+$-256 showcases the highest cycle count of 346 million cycles for key generation, interestingly lower than their cycles at security level III~\cite{sphincs}. For signing operations, HPPK-(128,1,1,1) demonstrates efficient performance with 16,046 cycles, showcasing the fastest scheme. Dilithium 5 is the second fastest scheme with about 2.8 million cycles, over 100x slower than HPPK DS. SPHINCS$^+$-256 exhibits the highest cycle count of almost 4.5 billion cycles. Falcon1024 is the second slowest scheme with over 22 million cycles. Verification sees HPPK-(128,1,1,1) maintaining efficiency with 22,285 cycles, outperforming Dilithium 5 and Falcon1024. SPHINCS$^+$-256 requires a substantial number of cycles due to its hash-based nature.

In summary, the performance of HPPK DS, particularly in signing and verification operations, is highly competitive across different security levels when compared to established NIST standardized schemes. The efficient use of cycles in HPPK DS, especially in scenarios with varying hash sizes, makes it a promising candidate for post-quantum cryptographic applications.

\begin{table}[htbp]
\caption{Comparison of signing and verification performance for HPPK DS is illustrated  with NIST standardized schemes. It should be noticed that the performance of HPPK DS is given based on sizes of hash algorithms: 32 bytes for level I, 48 bytes for level III, and 64 bytes for level V comparing with performance of NIST standardized algorithms using 32-byte hash-code. The performance in this table is from the HPPK DS configuration with $n=\lambda=m=1$ and the Barrett parameter $K=L+64$ bits. Performance data for Falcon and Dilithium without citations are directly computed from the NIST SUPERCOP tool.}
\begin{center}
\begin{tabular}{cccc}
\hline
\textbf{Crypto}&\multicolumn{3}{c}{\textbf{Performance (Cycles)}} \\
\cline{2-4} 
\textbf{system}& \textbf{\textit{KeyGen }}  & \textbf{\textit{Singing }}& \textbf{\textit{Verifying}}\\
\hline
    &\multicolumn{3}{c}{\textbf{Security Level I}} \\
    Dilithium 2~\cite{dilithium}$^1$ & 300,751 & 1,355,434 & 327,632 \\
Falcon512 & 38,194,993	& 10,303,471 & 68,621  \\    
SPHINCS$^+$-128s~\cite{sphincs}$^2$&358,061,994&2,721,595,944&2,712,044\\
HPPK-(64,1,1,1) & 25,696 & 12,510 & 18,349  \\ \\
   &\multicolumn{3}{c}{\textbf{Security Level III}} \\ 
Dilithium 3~\cite{dilithium}$^1$ & 544,232 	&2,348,703 & 522,267  \\
Dilithium 3     & 323,071	&1,418,393 & 313,271  \\
SPHINCS$^+$-192s~\cite{sphincs}$^2$&524,116,024&5,012,149,284&4,333,066\\
HPPK-(64,1,1,1) & 35,313 & 14,382 & 21,145  \\ \\
   &\multicolumn{3}{c}{\textbf{Security Level V}} \\
Dilithium 5~\cite{dilithium}$^1$ & 819,475	& 2,856,803 & 871,609  \\
Dilithium 5     & 454,296 & 1,479,623 & 483,62 \\
Falcon1024 & 101,629,055	& 22,423,017 & 138,671   \\
SPHINCS$^+$-256~\cite{sphincs}$^2$&346,844,762 & 4,499,800,456 & 6,060,438 \\
HPPK-(128,1,1,1) & 42,355 & 16,046 & 22,285  \\
\hline
\end{tabular}
\label{tab:DS_sign}
\end{center}
\footnotesize{$^1$ Average performance data are taken from their submission specification~\cite{dilithium}}\\
\footnotesize{$^2$ Performance data are taken from their NIST submission with hash SHA-256-simple in Table 4, using a single core of a 3.1 GHz Intel Xeon E3-1220 CPU (Haswell).}
\end{table}

\section{Conclusion}
\label{sec:conc}

This study provides a comprehensive evaluation of two innovative cryptographic schemes, HPPK KEM and HPPK DS, designed for post-quantum cryptographic applications. Through extensive benchmarking and comparisons with NIST-standardized algorithms, we have highlighted the key features and advantages of these schemes across various security levels.

For HPPK KEM, our analysis reveals a well-balanced combination of security and efficiency. The scheme's adaptability to different security requirements is evident in its key sizes, ciphertext sizes, and overall performance. The introduction of multiple noise variables adds a dynamic element to the encapsulation process, ensuring randomized operations even for the same secret. The efficient key generation, encapsulation, and decapsulation operations, particularly at higher security levels, position HPPK KEM as a promising solution for secure communication in a post-quantum era.

Regarding HPPK DS, our evaluation highlights its superiority in terms of compact key sizes and signature sizes across various security levels. The scheme demonstrates notable efficiency in key generation, signing, and verification operations. The innovative use of hidden rings, coupled with considerations for hash algorithm sizes, contributes to the compactness and efficiency of HPPK DS. Its competitive performance, especially in signing and verification, establishes it as a robust option for applications requiring secure and efficient digital signatures.

The comparative analysis with NIST-standardized algorithms, including Dilithium, Falcon, and SPHINCS$^+$, underscores the competitive nature of HPPK KEM and HPPK DS. These schemes outperform established algorithms in various performance metrics, showcasing their potential for practical deployment in real-world scenarios. Their robust security in homomorphic symmetric encryption, efficient performance, and adaptability to different security levels make them compelling choices for securing digital communication in the face of evolving cryptographic challenges. Future research directions may explore optimizations, conduct further security analyses, and investigate potential applications in emerging technologies.


\begin{backmatter}

\section*{Declarations}
\section*{Availability of data and materials}
All the data and materials generated are included in this manuscript. 
\section*{Competing interests}
The authors declare that they have no competing interests.

\section*{Author's contributions}
R.K. took the lead in drafting the manuscript and conducted the benchmark analysis. M.P., D.L., and B.T. collaborated to implement both Key Encapsulation Mechanism (KEM) and Digital Signature (DS) using the Supercop tool, generating comprehensive benchmarking results. All authors actively participated in the review process to ensure the quality and accuracy of the manuscript. 
\section*{Acknowledgements}
The authors express their gratitude to Xijian Zhu for his valuable assistance in reviewing the implementation codes. .


\bibliographystyle{bmc-mathphys} 
\bibliography{my}      

\end{backmatter}
\end{document}